# Unveiling Chirality: Exploring Nature's Blueprint for Engineering Nanostructured Materials

Alexa Guglielmelli[1], Liliana Valente[1], Giovanna Palermo[1], Giuseppe Strangi[1,2*]

[1]Department of Physics, NLHT-Lab, University of Calabria and CNR-NANOTEC, Institute of Nanotechnology, 87036 Rende, Italy.
[2]Department of Physics, Case Western Reserve University, 2076 Adelbert Rd, Cleveland, Ohio 44106, USA.

*Corresponding author(s). E-mail(s): giuseppe.strangi@fis.unical.it;

**Abstract**

Chirality, the property of asymmetry, is of great importance in biological and physical phenomena. This prospective offers an overview of the emerging field of chiral bioinspired plasmonics and metamaterials, aiming to uncover nature's blueprint for engineering nanostructured materials. These materials possess unique chiral structures, resulting in fascinating optical properties and finding applications in sensing, photonics, and catalysis. The first part of the prospective focuses on the design and fabrication of chiral metamaterials that mimic intricate structures found in biological systems. By employing self-assembly and nanofabrication techniques, researchers have achieved remarkable control over the response to light, opening up new avenues for manipulating light and controlling polarization. Chiral metamaterials hold significant promise for sensing applications, as they can selectively interact with chiral molecules, allowing for highly sensitive detection and identification. The second part delves into the field of plasmonics nanostructures, which mediate enantioselective recognition through optical chirality enhancement. Plasmonic nanostructures, capable of confining and manipulating light at the nanoscale, offer a platform for amplifying and controlling chirality-related phenomena. Integrating plasmonic nanostructures with chiral molecules presents unprecedented opportunities for chiral sensing, enantioselective catalysis, and optoelectronic devices. By combining the principles of chiral bioinspired plasmonics and metamaterials, researchers are poised to unlock new frontiers in designing and engineering nanostructured materials with tailored chiroptical properties.

# 1 Introduction

Chirality is a fundamental concept in the study of the natural world, referring to the lack of symmetry between an object and its mirror image. This concept is applicable at all scales, from the subatomic to the galactic, and holds great significance in various scientific fields, such as physics, chemistry, and medicine.[1] At the molecular level, the spatial asymmetry between enantiomers, which are mirror-image pairs of molecules, results in notable disparities in their chemical and physical properties. These distinctions play a pivotal role in numerous biological processes, including protein function, cell communication, and overall organism health. Additionally, chiral molecules are commonly found in pharmaceutical and agrochemical products, underscoring the importance of distinguishing between their enantiomers to ensure safe and effective usage.

However, current methods for detecting and separating chiral compounds are inefficient and costly, leading to the synthesis of many products as racemic mixtures, which can reduce their efficacy and cause harmful side effects.

Recent advancements in nanophotonics offer a promising avenue for the development of highly sensitive and efficient techniques in chiral detection, separation, and chiral signal enhancement. In this article, we explore the utilization of chiral symmetry in designing novel functional materials inspired by the natural chirality observed in biological systems. We also delve into the profound impact of chiral light-matter interactions, enabling crucial functions like recognition, separation, and activation. In particular, the emergence of new chiral plasmonic metasurfaces presents an exciting opportunity to revolutionize optical biosensors, introducing innovative methodologies to enhance their performance.

The exploitation of chiral light-chiral matter interaction can be considered a new paradigm for the next generation of optical biosensors to detect, characterize, and separate enantiomers down to the single molecule level, using design rules that can enable high-yield enantioselective interaction and selective manipulation. We also highlight future challenges and opportunities for this field.

# 2 The importance of structural and functional bio-chirality

Chirality plays a predominant role in the biological world, starting from amino acids that are building blocks of more complex chiral macromolecules. The majority of chiroptics' biological applications split into two primary groups: (1) studies of how chiral features affect fundamental biological processes, and (2) selection of chiral biomolecules' enantiomers as biomarkers for diagnosis and prognosis of the most aggressive and widespread diseases. The bio-category of nucleic acids is also governed by structural and functional chirality. The different types of DNA found in nature are chiral and have significant implications for DNA's functional interactions with proteins and drug molecules. Chirality introduces asymmetry when DNA is under torsional stress. Negative torques decrease helicity, and positive torques increase stability against strand separation.[2] Different points on the DNA chain are clamped together by proteins to prevent torsional relaxation.[3] These external stresses have an impact on the DNA chain's larger-scale structure as well as how the genetic material is stored inside the cell, in addition to its local characteristics. Many features of DNA torsional biophysics can be investigated using non-invasive chiroptical techniques. Circular dichroism (CD) was used to discover the left-handed helix structure of the so-called Z-DNA.[4, 5] Misregulation of Z-DNA is involved in pathological conditions [6] like Alzheimer's, [7] and autoimmune disorder .[8] Additionally, CD analysis can pro- vide insight into drug therapies that modify topoisomerase activity. These enzymes are significant clinical targets for antibiotics and cancer cures because they are necessary for regular DNA replication and transcription. [9] These inhibitory pharmaceuticals intercalate DNA by inserting themselves in between base pairs and engaging with the enzyme to prevent DNA reattachment, this alters the DNA's local chiral structure in a manner that from CD signal can be detect.[10]

Intrinsic molecular chirality propagates across multiple levels affecting large-scale functionality, infact living organisms exhibit chirality not only at the molecular level but also at a macroscopic scale (see Figure 1).[11]

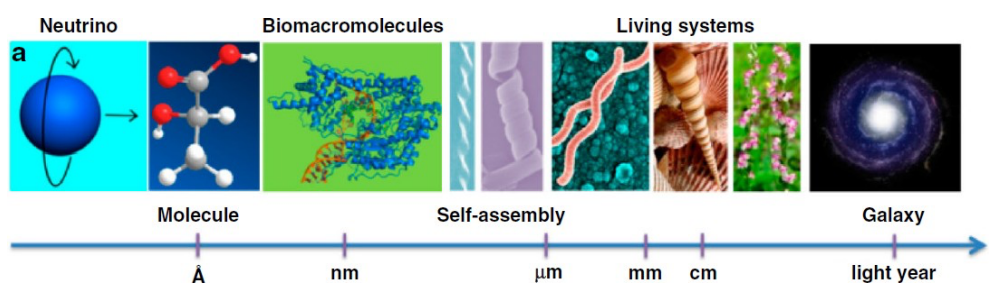

**Fig. 1** Multiscale chiral architectures in a variety of hierarchical levels, ranging from atomic, molecular to supramolecular, macroscopic, and galactic scales. Reprinted with permission from [11]. Copyright 2015, American Chemical Society.

Asymmetries exist from the molecular to the organismal level and can affect tissue and behavioral levels.[12] Correct development of left-right (L/R) asymmetry in embryonic development is essential for proper organ location and morphogenesis.[13] Macroscopic asymmetries in living organisms may be linked to biomolecular chirality, and the fruit fly is a fascinating case study.[14] The chiral looping for tubular organs, which involves a right-handed loop, is controlled by the motor protein Myo1D. Myo1D is able to exert a chiral rotational force on actin and plays a role in the chiral organization at all higher levels. By gaining insights into these processes, we can derive inspiration for the creation of engineered designs that incorporate chirality across multiple scales, thereby leading to functional nanomaterials. Moreover, the ability to modify the concentration of the molecules involved provides a means to achieve tunability in both the chiral structure and chiroptical response.

The close link between biomolecular chirality and functionality has promising medical applications. Specific chiral molecules can be used as biomarkers for disease diagnosis and prognosis, and these biomarkers are ideal targets for highly sensitive chiroptical techniques that can discriminate among stereoisomer. Researchers have investigated different groups of chiral molecules as potential predictive biomarkers, including byproducts of oxidative stress,[15] metabolites,[16, 17] and amino acids.[18, 19] Isoprostane regioisomers, produced when fatty acids oxidatively degrade under the influence of free radicals, are candidate biomarkers for oxidative stress-caused diseases and pathological development of disease like Alzheimer's.[15] Recently growing attention has been devoted to the biological significance of chiral metabolites, such as the crucial function of D-Asp in hormone synthesis and vertebrate aging,[20, 21] the critical roles of D-Ser in kidney and brain disease,[22] the possible function of D-Gln as a serum biomarker in hepatocellular carcinoma,[23] and the close relationship between gastric cancer and chiral amino-containing metabolites.[17] In addition, enzyme mutations can result in the creation of novel chiral metabolites that aid in the initiation and development of cancer. Isocitrate dehydrogenase 1 (IDH1) and 2 (IDH2) genes characterize both low-grade and high-grade glioma patients, and their activity results in the elevated production of the d-enantiomer of the metabolite 2-hydroxyglutarate.[24] Compared to their l-counterparts, the d-forms of amino acids (d-AAs) are found in living organisms at relatively low concentrations. Researchers are working to identify particular d-AAs in the human body as biomarkers for neurological diseases like schizophrenia and amyotrophic lateral sclerosis as well as age-related conditions like atherosclerosis and cataracts.[25] In the early phases of dementia, astrocytes' sugar metabolism is associated with the d-serine (d-Ser), suggesting that the enantiomeric content of d-AAs represent an appropriate chiral candidate biomarker for this disease diagnosis.[26] Furthermore, it was shown that the quantities of d-Ser in the spinal cord of amyotrophic lateral sclerosis model mice it is higher compared to control mice, and the change can be linked to the development of the illness.[27] As human chronic kidney disease patients' conditions intensified, the levels of numerous d-AAs, including serine, proline, and alanine, increased noticeably.[28] All these findings suggest that d-AAs may serve as novel biomarkers for different abnormal physiological conditions and disease, like renal failure and can also be detected to monitor the cancer cell proliferation and infections caused by different bacteria strains.[29, 30]

## 3 Selectivity in Adsorption on Biointerfaces and Composite Chiral Surfaces

Chiral recognition and separation methods for bio-enantiomers are in high demand for medical-pharmaceutical applications and can be challenging due to their similar physical and chemical properties. Chiral enantiomers possess identical physical and chemical properties except for their interaction with other enantiomeric compounds and chiral light. The profound implications that chirality could have on our health strongly came to the fore after the discovery of catastrophic adverse effects of the racemic drug thalidomide, which was prescribed to treat morning sickness during pregnancy. While the D-Thalidomide was safe, with a sedative effect, its mirror image turned out to be teratogenic, causing severe birth defects. [31]

Chiral discrimination of enantiomers is critical in various fields, including pharmaceuticals, chemistry, and biology. However, traditional methods for chiral discrimination, such as High-performance Liquid Chromatography (HPLC), Gas chromatography, and Capillary electrophoresis, can be expensive, time-consuming, and difficult to adapt for point-of-care devices. Moreover, these methods require exclusive chiral selectors and have poor recognition resolution, low sensitivity for different chiral targets, and are not label-free. In contrast, chiroptical techniques, including Circular Dichroism Spectroscopy, rely on the inherent optical activities of chiral molecules. However, these techniques also suffer from sensitivity limitations, particularly for small molecular mass biomolecules since the CD signal for such chiral compounds is usually weak. Therefore, there is a need for new methods for chiral discrimination that can overcome these limitations and provide high sensitivity and specificity for chiral molecules, especially for small molecular mass biomolecules.

The behavior of chiral molecules and drugs in relation to their specific structures has been extensively studied, particularly in terms of their selective binding to functional proteins such as receptors and enzymes. However, there has been limited consideration given to the potential of lipid membranes in detecting the enantiomeric differences of active biomolecules and handedness specific responses [32]. One intriguing application that has been explored involves incorporating molecular micelles into capillary electrophoresis separations as chiral selectors [33]. These separation process can be challenging because it relay on small differences in binding free energies between analyte enantiomers and the chiral selector.[33] Additional unconvincing approaches involve the use of polymeric solid membrane or membrane imprinted towards one specific enantiomer in a continuous process.[34] Lipid membranes have proved to be a valid biological platform for carrying out chiral recognition of small molecules and drugs, like Tryptophan (Trp) and Ibuprofen (IBU), as shown in Figures 2a,b. [35–37] Compared to other techniques, lipid membranes offer the enormous advantage of being highly selective and specific simply by changing the percentages and lipid composition of the membranes used. [38–41] Additionally, using model lipid membranes can help gain insight into the stereochemistry aspects of interactions between chiral molecules and cell membranes. [42] In this regard, one of the possible explanations for the different effects of drug enantiomers is the presence of chiral carbon atoms in phospholipids that induce a stereostructure-specific membrane interaction.[43] The chiral recognition mechanism relies on the formation of multiple interactions at the polar/apolar lipid interface, as it is schematically illustrated in Figures 2a,b.[36, 37] These findings also suggest the significance of the membrane properties, i.e., surface charge, membrane fluidity, and membrane polarity, in understanding the enantioselective adsorption behaviors of chiral AAs on the self-assembled liposomes. In this context, our recent work showed the interaction of the two chiral forms of Tryptophan (L-/D-Trp) with the polar/apolar interfaces of two different model membranes made of DPPC (neutral) or DPPG (negatively charged) lipids. The weak enantiomeric specificity found resulted to be modulated by the nature of the lipid polar head (i.e., phosphatidylcholine versus phosphatidylglycerol) and by the physical state (i.e., gel versus liquid-crystalline) of the lipid membranes.[44] A recent study exploited a self-assembled DPPC lipid membrane as a coating for porous polymer particles to selectively adsorbs L-Trp, showing a high rate of enantioselectivity below the phase transition temperature of DPPC.[41] These achievements could help to conceive tailored lipid membranes and lipid membrane–polymer composite for achieving selective discrimination of chiral biomolecules. To be used for biomedical and pharmaceutical advanced applications the membrane composition can be suitably designed to maximize the enantiospecificity with different chiral biomarkers or drugs.

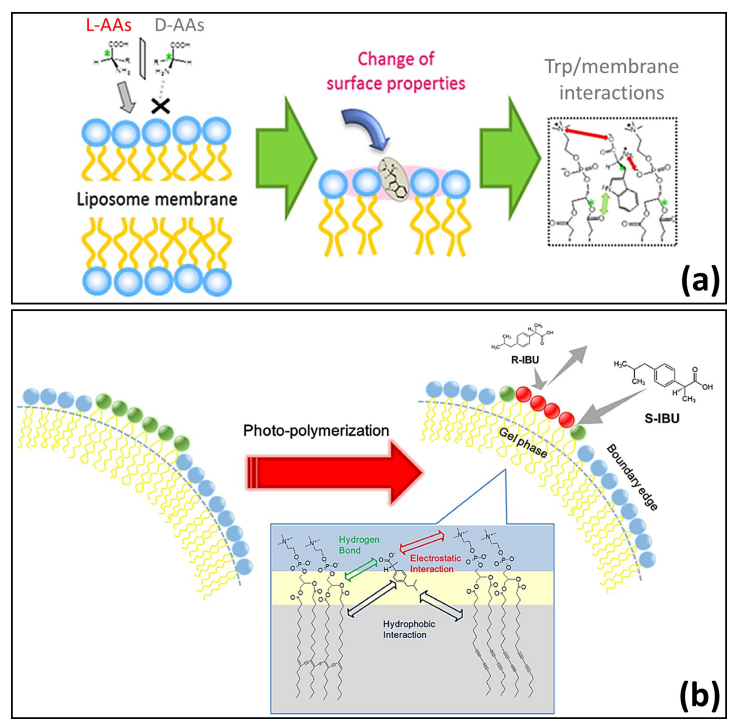

**Fig. 2** Schematic illustrations (**a**) of the mechanism of stereoselective interaction of amino acids on DPPC liposomal membrane. Adapted with permission from [36]. Copyright 2015, American Chemical Society. (**b**) of the chiral recognition of ibuprofen mechanism by polymerized diacetylenic liposomes. Reprinted with permission from [37]. Copyright 2017, American Chemical Society.

Understanding the underlying processes involved in the stereo-specific interactions between chiral enantiomorphous nanoarchitecture and biosystems is of utmost importance for two key reasons. Firstly, it lays the foundation for the development of novel, high-performance biocompatible materials. By comprehending how chiral structures interact with biosystems, researchers can design and engineer materials with enhanced properties that can be used in various applications. This has the potential to revolutionize fields such as medicine, biotechnology, and materials science. Secondly, gaining insights into the high selectivity mechanisms between chiral biosystems expands our knowledge of the intricate workings of biological systems. By studying how chiral entities interact with biosystems, we can unravel the mechanisms behind their selective behaviors. This understanding not only deepens our comprehension of fundamental biological processes but also paves the way for advancements in drug discovery, diagnostics, and therapeutic interventions. It was shown that immune cells display chiral-dependent behaviors when exposed to various enantiomers-functionalized surfaces. [45] In particular, to investigate the stereospecific behavior of cells on chiral-like surfaces, immune cells were tested on a gold-sputtered surface functionalized with N-isobutyryl-L(D)-cysteine (NIBC) enantiomers, used as an extracellular matrix substrate. The results show that cells adhered on the two enantiomeric surfaces displayed enantio-differences in quantity and shape. In particular cells on the D-NIBC Au surface exhibited lower density and smaller spreading area compared to the one obtained on L-NIBC Au surface, as shown in Figures 3a, b. The induced surface chirality also had an impact on the cells' shapes; on the D surface, cells maintain their separation and have a roughly rounded morphology, whereas, on the L surface, the majority of the cells have a more complex morphology exhibiting a highly spreading pattern. These results show that the D-NIBC modification can decrease the cells' adhesion on the surface, while the L-NIBC has the opposite effect. A similar effect applies to different cell types, like Mesenchymal stem cells (MSCs). [46] It was shown how the surface molecular chirality affects the adhesion and differentiation of stem cells by using self-assembly monolayers of L- or D-cysteine (Cys) on a glass surface coated with gold. The fluorescence micrographs in Figures 3c, d show a higher MSCs density on the L-Cys Au surface and greater cells' spreading on the D-Cys Au surfaces. The cell adhesion was enhanced by the L surface, probably owing to the preferred adsorption of serum proteins. In addition, the inherent supramolecular chiral effects on dental pulp stem cells (DPSCs) spreading and differentiation were studied. [47] From Figures 3e,f it can be seen that left-handed nanofibers assembled from L-amino acid derivatives enhance cell spreading and proliferation, while right-handed nanofibers have the opposite effect, challenging the traditional belief that cell adhesion is usually improved by a fibrous morphology. These findings imply that supramolecular helical handedness significantly controls cell behavior and in addition that chirality influences cell differentiation. Again a possible explanation of the promotion of cell adhesion on the L-surface comes from the preferential protein adsorption on that surface. Another interesting selective cells' proliferation study was performed on monolayer plasmonic chiral AuNPs films modified with L- or D-penicillamine (L-/D-Pen).[48] The L-Pen-NP films promote cell proliferation, while the D-Pen-NP films have the opposite impact. In fact, as can be deduced by looking at Figures 3g,h cells on L-Pen-NP film cover a comparatively larger area than those on D-Pen-NP film. In addition, the presence of a chiral plasmonic film allows the study of the photo-thermal induced cell detachment using circularly polarized light. The investigation and implementation of these effects will accelerate cell culture and engineering, which includes the use of stem cells for therapeutic purposes. In-depth research may influence the creation of artificial prosthetic devices that mimic biological or natural processes using biomaterials.

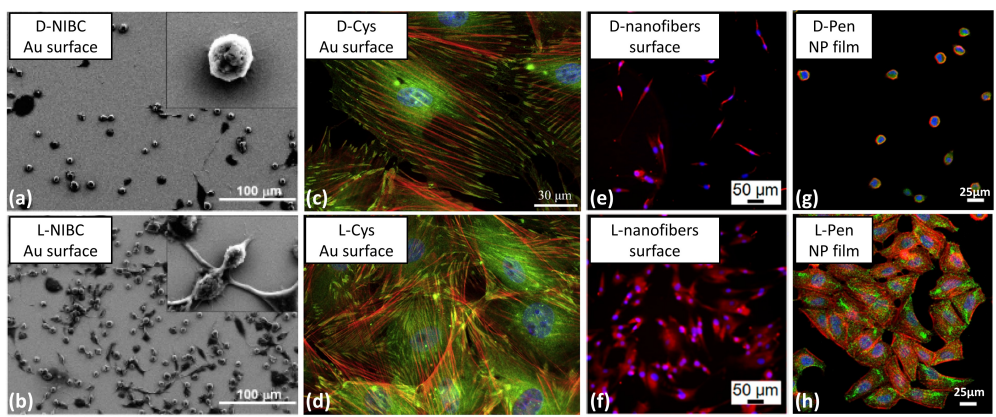

**Fig. 3** The interaction between cells and chiral surfaces. SEM images of immune cells adhesion on (**a**) D-NIBC-modified and (**b**) L-NIBC-modified Au sputtered surfaces after 24 h of cell incubation; Adapted with permission from [45]. Copyright 2007, American Chemical Society. Fluorescence micrographs of stem cells adhering on the (**c**) D- and (**d**) L-Cys modified Au surfaces after 4 days of culture; Adapted with permission from [46]. Copyright 2013, Elsevier Ltd. Fluorescence microscopy images of DPSCs on (**e**) Right- and (**f**) Left-handed nanofibers modified surfaces after 1 day culture; Adapted with permission from [47]. Copyright 2019, American Chemical Society. Confocal images of HeLa cells on (**g**) D- and (**h**) L-Pen-NP film. Reproduced under the terms of the CC-BY Creative Commons Attribution 4.0 International license (https://creativecommons.org/licenses/by/4.0).[48] Copyright 2017, Springer Nature.

## 4 Chiral Bioinspired Plasmonics and Metamaterials

In recent years, the investigation of chiral bioinspired plasmonics and metamaterials has gained significant importance, driven by their immense potential in various fields such as sensing, drug discovery, and nanotechnology. Chirality, a fundamental concept in nature, holds a crucial role in the intricate interactions between light and matter. Consequently, the development of bioinspired plasmonics and metamaterials has emerged as a highly promising avenue for designing chiral structures with customized properties. Inspired by the chirality found in biological systems, such as proteins and DNA, these structures mimic the geometry and function of natural chiral systems. This approach allows for the design of new materials with chiral properties, such as circular dichroism (CD) and optical activity, which are critical for chiral sensing and discrimination.

Recently, it has become popular to develop artificial chiral nanoparticles that perform some of the same tasks as native DNA-binding enzymes. Significant examples are chiral quantum dots, which can cut DNA or proteins, like tetrahedral cadmium telluride (CdTe) NPs functionalized with a L-cysteine for DNA cleavage applications.[49]

The completely DNA cleaveage in a specific position, could be triggered with circularly polarized light by producing reactive oxygen species (ROS). Another artificial enzyme, a chiral carbon dot, can intercalate DNA and imitate the action of an enzyme by cleaving only one strand of DNA, with production of ROS.[50] The emerging interaction between the chiral carbon dot and the DNA turned out to be stereoselective, with the D-form more efficient than the L-form. In this context using CPL and chiral nanoparticles could paved the way for gene editing technique enabling gene manipulation at the molecular level.

Achiral metallic nanoparticles can be efficiently combined with naturally occurring chiral biomarkers to give rise to intense CD signals. In particular amyloid fibrils have been used as chiral templates to confer an enantio-morphology to gold nanorods (AuNRs) (Figure 4a,b) and at the same time perform detection of biomarkers for amyloid diseases.[51] The AuNRs could adsorb onto protein fibrils with an helical-like structure but do not associate with monomeric proteins. The AUNRs with samples from Parkinson's disease patients produced strong CD signals that ranged in wavelength from 500 nm to 900 nm, due to the AuNRs 3D chiral arrangement dictated by the helical structure of the template (Figure 4c). In control experiments using healthy brain samples, only marginal CD signals were detected.

Another example of chiral assemblies of biomolecules and metallic NPs are the self-assembled gold nanohelices made of AuNPs of 10 nm diameter arranged around the surface of DNA origami (see Figure 4d).[52] Figure 4e and 4f show the expected CD signal for an helical arrangement of plasmonic NPs with a characteristic bisignate peak–dip shape, with the peak centered around the plasmon resonance frequency.

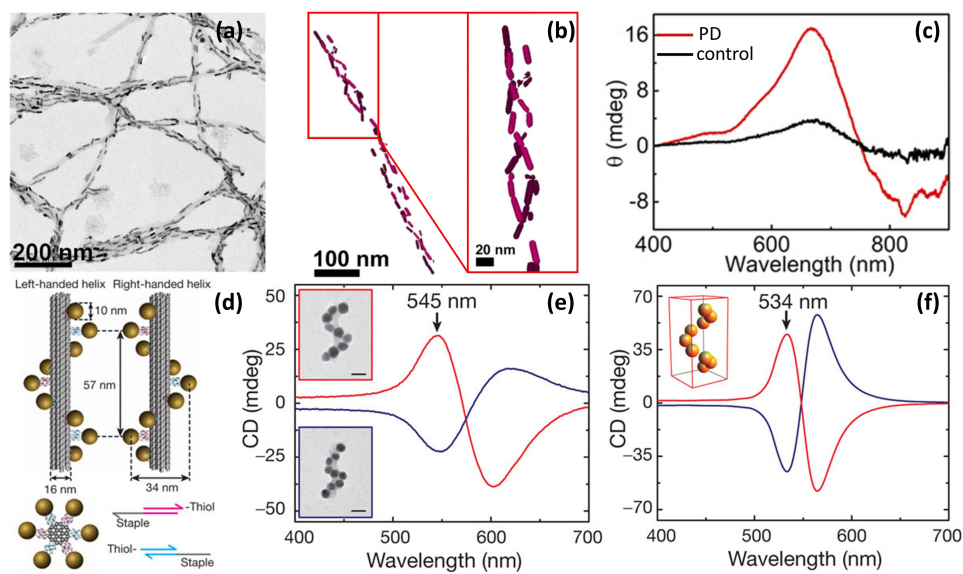

**Fig. 4** (**a**) TEM images of AuNRs (0.5 nM) in the presence of α-synuclein fibrils; (**b**) Cryo-TEM tomography reconstruction showing the 3D chiral arrangement of AuNRs on fibrils, along with a zoom of the red squared region of the assembly; (**c**) CD spectra of AuNRs after the addition of purified brain homogenates from healthy (control-black curve) and PD-affected (red curve) patients. Adapted with permission from [51]. Copyright 2018, National Academy of Sciences. (**d**) Left- and right-handed self-assembled nanohelices (diameter 34 nm, helical pitch 57 nm) formed by nine AuNPs each of diameter 10 nm that are attached to the surface of DNA origami. AuNPs are functionalized via thiol-modified DNA strands that are complementary to binding sites on DNA surface. (**e**) CD experimental and (**f**) theoretical spectra of Au left-handed (red curves) and right-handed (blue curves) nanohelices, along with the TEM images and models in the insets. Adapted with permission from [52]. Copyright 2012, Springer Nature Limited.

## 5 Plasmonics nanostructurers mediated enantioselective recognition via optical chirality enhancement

The investigation of optically active compounds, from tiny molecules with a single chiral center to supra-molecular assemblies, is accomplished via circular dichroism (CD) spectroscopy. CD spectroscopy can discriminate between enantiomers and determine the enantio-purity level of mixed analytes, by measuring the difference in the absorption of circularly polarized light (CPL) by chiral compounds. By locating spectral signatures indicative of structural patterns, CD spectroscopy can also be used to track the conformations of biomolecules such as peptides, proteins, and oligonucleotides. Bioanalytical uses for CD spectroscopy include assessing conformational changes in aptamers brought on by target engagement, determining the environmental stability of oligonucleotides, and understanding protein aggregation linked to illness development and progression. The moderate differential CPL absorption by chiral compounds as compared to the overall absorption of unpolarized light, however, limits the sensitivity of CD spectroscopy. To overcome this restriction, subwavelength nanophotonic structures can be employed to preferentially boost the rate of light absorption by enantiomers over their mirror image isomers. These structures improve the intensity and chirality of evanescent fields. Through near-field and/or far-field interaction with the plasmonic resonance of the metal nanoparticles, the chiroptical characteristics of enantiomers in molecular plasmonic assemblies can be significantly improved and controlled.

Plasmonic chiral metamaterials and superchiral fields have been proposed to enhance enantioselective precision at low molar sensitivity. Hendry et al. [53] reported the first chiral biosensing technology using a metasurface, and similar results were achieved by exploiting the enhanced sensitivity of superchiral evanescent fields to the chiral structure of different biomolecules. Plasmonic metamaterials widen and shift the CD band of molecules in the visible region of the electromagnetic spectrum, providing a significant advantage in their detection. Despite the ongoing challenges associated with chiral discrimination of enantiomers, recent progress in the fields of chiral metamaterials and microfluidic platforms offers promising solutions.[54, 55] These advancements hold the potential to enhance the sensitivity and efficiency of chiral sensing techniques, enabling rapid detection of both achiral and chiral biomolecules. This breakthrough has significant implications for various applications requiring precise identification and characterization of enantiomers in a timely manner. Esposito et al. demonstrated the sensing capabilities of an ordered array of 3D core-shell chiral nano-helices fabricated by means of the focused ion beam induced deposition (FIBID) by providing a refractive index sensitivity of 800 nm/RIU, which is among the highest values obtained in LSPR based sensors, thanks to the large binding area (Figure 5a,b). The platform has been used to detect the presence of TDP-43 protein in human serum, a protein clinically relevant for neurodegenerative diseases, in a very low concentration range (1 pM - 10 fM) - Figure 5c,d.[56]

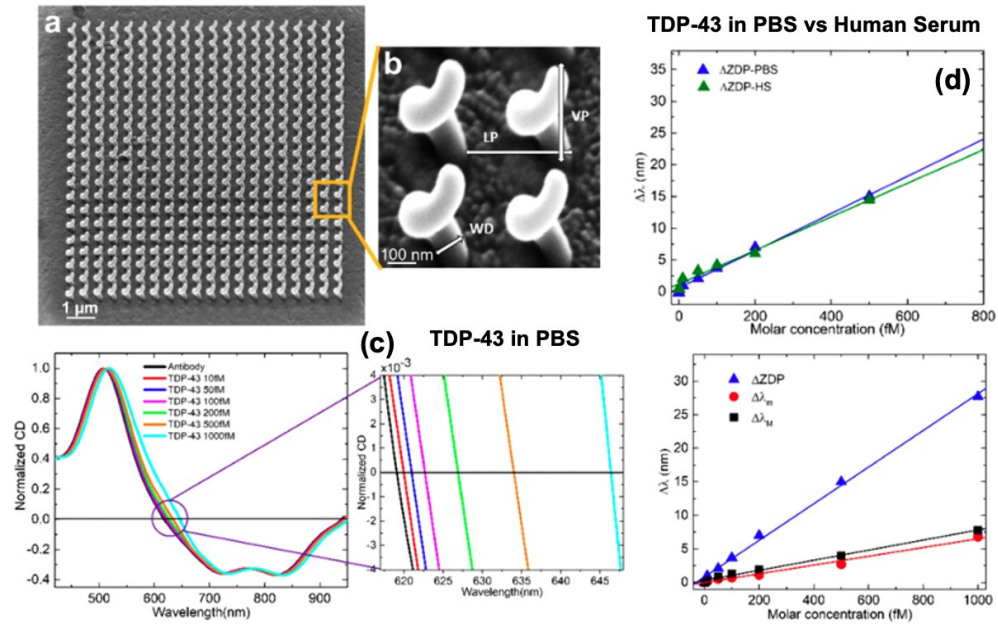

**Fig. 5** (**a**) SEM images of the 3D array of Pt-nanohelices array. (**b**) Details of the geometrical parameters of the nano-helix (LP: lateral period, VP: vertical period, WD: wire diameter). (**c**) CD spectra of the metasurface after the deposition of different concentration of TDP-43; a magnification of the zero dichroism point around 600 nm; the wavelength shift of the maximum ($\lambda_M$), zero ($\lambda_{ZDP}$) ,and minimum ($\lambda_m$) as a function of the molar concentration. (**d**) Wavelength shift of the $\lambda_{ZDP}$ as a function of the molar concentration of TDP-43 in PBS and human serum. Adapted with permission from [56]. Copyright 2021, American Chemical Society.

Tang and Cohen, quite recently, introduced the optical chirality (*C*), a fundamental property of the electromagnetic fields, that quantify the intrinsic chirality of the electromagnetic waves.[57] In particular, they hypothesized that the inherently weak CD response of chiral molecules could be enhanced by tailoring the *C* of the interacting fields. This can be obtained by decrease the electric energy density [58] or by the incorporation of plasmonic nanostructures. In this context, Kadodwala et al. proposed a "superchiral spectroscopy" technique, demonstrating that "shuriken" nanostructures made of Au had evanescent waves that boost *C*.

Enhancing *C* through manipulation of near fields is one factor that can lead to higher sensitivity CD, since the CD signal of a molecule is related to the differential absorption of right- and left-CPL that is proportional to the optical chirality.[59]

Kadodwala et al. describe how to use this phenomenon to detect with high sensitivity changes in protein structure at the picogram level thanks to the enhancement of *C* associated to the evanescent field related to the LSPR.[60] This unlocked opportunities for numerous innovative designs of near-field factor *C* and enabled picogram detection of the chiral molecules immersed in the chiral near-field, that asymmetrically change the chiroptical response of the enantiomorphic structures.[61] The *C* enhancement can be achieved through different plasmonic chiral shapes, and also with completely symmetric nanostructures.[62, 63] For example, achiral plasmonic dimers showed plasmon-enhanced near fields resulting in a CD boosting up to three orders of magnitude.[64] These findings underlined that achiral nanostructures locally enhance optical chirality amplifying the molecule's chiral response. [65] The sensing of chiral molecules using chiral metasurfaces is still in its early stages of exploration. Currently, the main bottleneck lies in the fabrication methods, specifically Focused Ion Beam Induced Deposition (FIBID) and electron beam lithography (EBL). These techniques are costly and result in small sensing regions, limiting their scalability and practicality. Additionally, there is a need for further research on the design and optimization of chiral metasurfaces, particularly in enhancing the chiral near fields. This aspect will play a significant role in future studies. To overcome these challenges, alternative manufacturing techniques such as Nanoimprint Lithography show promise. By enabling large-scale production and cost reduction, Nanoimprint Lithography could revolutionize the development of chiral biosensors. Moreover, the incorporation of Artificial Intelligence (AI) techniques for optimization design will be crucial. AI can aid in efficiently exploring and optimizing the vast design space of chiral metasurfaces, leading to enhanced performance and sensitivity of chiral biosensors. Overall, the future of chiral biosensors lies in addressing the current limitations of fabrication methodologies, exploring alternative manufacturing techniques, and utilizing AI for design optimization. These advancements will pave the way for the widespread adoption and commercialization of chiral biosensors, opening up new opportunities in various fields such as pharmaceuticals, healthcare, and environmental monitoring.

## 6 Conclusion

In conclusion, the field of chiral bioinspired plasmonics and metamaterials presents exciting opportunities for engineering nanostructured materials with tailored chiroptical properties. Chirality, as a fundamental property of asymmetry, plays a pivotal role in various biological and physical phenomena. By mimicking nature's intricate structures, chiral metamaterials have demonstrated remarkable control over the chiroptical response, enabling novel approaches in light manipulation and polarization control. These materials hold great potential for sensing applications, allowing for selective interactions with chiral molecules and achieving ultrasensitive detection and identification. Moreover, the integration of plasmonic nanostructures with chiral molecules in the field of plasmonics nanostructures mediated enantioselective recognition via optical chirality enhancement has opened up unprecedented possibilities. Plasmonic nanostructures, capable of confining and manipulating light at the nanoscale, provide a platform to amplify and control chirality-related phenomena. This integration offers opportunities for chiral sensing, enantioselective catalysis, and the development of optoelectronic devices. However, there are challenges to overcome. Fabrication methodologies, such as FIBID and EBL, currently limit the scalability and cost-effectiveness of chiral metasurfaces. To address this, alternative manufacturing techniques such as Nanoimprint Lithography show promise in enabling large-scale production and reducing costs. Furthermore, leveraging Artificial Intelligence (AI) techniques for optimization design can efficiently explore the design space of chiral metasurfaces and enhance their performance. As these challenges are tackled, the field of chiral bioinspired plasmonics and metamaterials is poised to unlock new frontiers in engineering nanostructured materials with tailored chiroptical properties. These advancements have the potential to revolutionize diverse fields, from energy harvesting to pharmaceutical development, bridging the gap between fundamental research and practical applications. With continued interdisciplinary collaboration and innovation, the future of chiral biosensors holds tremendous potential for rapid and sensitive detection of chiral and achiral biomolecules, ultimately transforming various industries and benefiting society as a whole.

**Acknowledgments.** Authors acknowledge financial support from the "NLHT - Nanoscience Laboratory for Human Technologies" (POR Calabria FESR-FSE 14/20).

**Conflict of Interest.** On behalf of all authors, the corresponding author states that there is no conflict of interest.